# Étendue and radiance conservation in transformation optics establish strict analytical bounds on field enhancement

**Mohammad Mehdi Sadeghi & Mustafa Sarısaman**

# Étendue and Radiance Conservation in Transformation Optics Establish Strict Analytical Bounds on Field Enhancement

Mohammad Mehdi Sadeghi [1,2,*], Mustafa Sarısaman[2]
[1] *Department of Physics, Jahrom University, Jahrom 74137-66171, Iran*
[2] *Department of Physics, Istanbul University, 34134, Vezneciler, Istanbul, Turkey*

**Corresponding Author:* sadeghi@istanbul.edu.tr
sadeghi@jahromu.ac.ir

**Abstract:** We establish an intuitive connection between transformation optics (TO) and the classical invariants of étendue and radiance. Through explicit application of the optical-metric formulation of TO, we demonstrate that any smooth, passive, impedance-matched transformation performs as a canonical (symplectic) mapping on optical phase space. Combined with Hamiltonian ray dynamics, this implies that Liouville's theorem applies as well to transformation-optical media, enforcing phase-space volume preservation and fundamental constraints on radiance and étendue under passive mappings. To our knowledge, this explicit phase-space formulation has not been systematically developed within the transformation-optics framework. We derive strict analytical bounds on achievable field enhancement. In particular, we show that the maximum average intensity attainable in any passive TO concentrator is limited solely by the geometric area-compression ratio of the underlying coordinate transformation, independent of the specific material realization. We apply this framework to zero-index media, optical-null media, and illusion devices and find the same rules.

Consequently, our findings demonstrate that TO redistributes optical intensity without increasing radiance, consistent with a Liouville-type constraint. This result provides a consistent, metric-based explanation for fundamental concentration limits in passive, impedance-matched metamaterials.

## 1. Introduction

Transformation optics (TO) provides a geometric framework for manipulating wave propagation by conceptualizing coordinate transformations as effective material responses [1–3]. Anisotropic and inhomogeneous permittivity and permeability tensors encode a desired mapping between a "virtual" space where Maxwell's equations are simple and a "physical" space where the fields are implemented within this approach.[1–3] This perspective has led to the development of various devices, including electromagnetic cloaks, field concentrators, illusion media, and optical analogues of gravitational metrics.[1–7] TO-based devices commonly operate by compressing one region of space and expanding another. This is especially true for extreme-index platforms like zero-index and optical-null media, which can strongly confine and focus electromagnetic fields [4,5,8–10,20].

Such devices are frequently interpreted as if they could produce extremely large field concentrations: when a finite input aperture is mapped, via a coordinate transformation, into a much smaller physical region, the resulting geometrical compression appears to suggest unbounded intensity growth.[4–6,9] At the same time, classical radiometry and nonimaging optics have long emphasized that there are strict limits to how much light can be concentrated by any passive optical system. These limits are expressed in terms of radiance and étendue: radiance (brightness) is the power per unit projected area and per unit solid angle, while étendue, a French term literally meaning “extent”, is the product of area and solid angle that quantifies how spread a beam is in phase space [11–13,15]. In lossless, passive systems, radiance cannot increase and étendue cannot decrease; this is the basis of the classical brightness theorems and the principle of conservation of etendue [11–13]. In wave language, these invariants are interpreted as the consequences of Liouville's theorem when it is used on the Hamiltonian flow of rays in phase space.[10,15,16] The evolution of rays (x,k) is governed by a Hamiltonian that encodes the local dispersion relation, and the corresponding flow is symplectic, preserving phase–space volume $d^3 x, d^3 k$. In anisotropic media, the invariants can be articulated with greater precision; however, the fundamental conclusion persists: in passive, linear, geometrical optics, radiance remains conserved along rays, and étendue serves as a strict upper limit on the attainable concentration of power.[11,12,14,15] Although these concepts form the basis of many concentration limits in imaging, solar energy, and wave scattering, they are still not often explicitly addressed in TO and optical metrics literature.[3, 11, 14] This gap has both practical and theoretical effects. In practice, devices such as TO-based concentrators, cloaks, and illusion devices are frequently constructed and analyzed solely at the level of fields or scattering cross-sections, without making an explicit connection to radiance or étendue.[4–7,9,10,17–19]

Extreme-index metamaterials like zero-index and optical-null media can make the local field much stronger. This makes us wonder if TO-based designs may extend around the limits of classical brightness.[4,5,8–10,20] From a conceptual standpoint, there is currently no overarching declaration in the TO formalism that defines what is rigorously invariant and what is fundamentally permissible in terms of field enhancement under passive coordinate transformations. In particular, the distinction between intensity (local power per unit area) and radiance (power per unit area per unit solid angle) is often blurred, leading to confusion between legitimate intensity amplification and forbidden increases of brightness [11-13]. In this work, we establish a definitive relationship between TO and the classical invariants of radiance and étendue. In the geometrical-optics limit, we demonstrate that the effective optical metric generated by a TO mapping induces a Hamiltonian ray system whose phase–space evolution is symplectic.[1–3,16] As a result, the TO mapping is a canonical transformation in (x,k)-space, and the phase–space volume element, and hence étendue, is strictly conserved in any passive, lossless TO device.[11,12,14–16] Utilizing this framework, we extract strict analytical bounds on field enhancement: for a given input radiance and numerical aperture, the local intensity in the physical device cannot exceed a bound determined by the optical metric and the Jacobian of the coordinate transformation. In other words, TO can strongly redistribute intensity, but it cannot increase radiance; any apparent "super-concentration" is constrained by étendue conservation.[11–14] Our study improves and specializes recent scattering concentration bounds and brightness theorems that address general passive linear systems, by emphasizing the significant subclass of media produced by TO.[14] Within this subclass we obtain closed-form expressions for the radiance-invariant phase–space measure, clarify the role of anisotropy in the optical metric, and make explicit the assumptions under which radiance and étendue are conserved. Next, we apply the general bounds to several canonical TO devices, such as ideal concentrators, zero-index and optical-null-based structures, and illusion media. This lets us figure out how close practical designs come to saturating the theoretical limits, and to identify regimes where diffraction, loss, or gain are required to apparently "beat" the geometrical bound.[4–7,9,10,17–20]

This work aims to focus on passive, linear, time-invariant systems in the geometrical-optics regime, where a Hamiltonian description of rays is valid and absorption or scattering can only decrease radiance [11–13,16]. Under these assumptions, our findings can be interpreted as a Liouville-type theorem for TO: TO inherits, and does not override, the fundamental conservation of étendue and radiance.[11,12,14–16] The rest of the paper is organized as follows. In Section 2, we examine the TO formalism and present the optical metric and Hamiltonian ray equations. In Section 3, we construct the conserved phase-space measure and establish radiance and étendue within this metric framework, resulting in analytical upper limits on intensity and field enhancement. In Section 4, we use these constraints on some TO devices and talk about how near they are to the fundamental limits. In Section 5, we end with a conclusion.

## 2. The optical metric and TO formalism

In this section we study only the minimal TO and ray-dynamical ingredients needed for the phase-space invariants and intensity bounds derived in Sec. 3. Our purpose here is not to review textbook ray optics, but to make explicit one key structural point for smooth, passive, impedance-matched TO mapping induces a canonical (symplectic) transformation in optical phase space. This statement is the precise bridge between TO and the radiance/étendue constraints developed later.

### 2.1. Coordinate transformation and impedance-matched TO media

Consider a smooth bijective mapping between a virtual space $x^i$ and a physical space $x'^i = f^i(x)$, with Jacobian

$\Lambda^i{}_j = \frac{\partial x'^i}{\partial x^j}, J = \det\Lambda$. If the virtual space is characterized by constitutive tensors $\varepsilon_v, \mu_v$, standard TO construction yields the effective medium in physical coordinates

$\varepsilon_p = \frac{\Lambda\, \varepsilon_v\, \Lambda^T}{J}, \mu_p = \frac{\Lambda\, \mu_v\, \Lambda^T}{J}$. In the common case $\varepsilon_v = \varepsilon_0 I$, $\mu_v = \mu_0 I$, the transformed medium is impedance-matched ($\varepsilon_p \propto \mu_p$) and purely passive provided the mapping is smooth and the material implementation is lossless. The geometric content of the mapping is encoded by J: compression ($|J| \ll 1$) and expansion ($|J| \gg 1$) of spatial elements will later be shown to be accompanied by reciprocal reshaping in the angular/momentum sector.

### 2.2. Optical-metric representation of the local dispersion

For impedance-matched TO media, it is convenient to represent wave propagation using an effective optical metric. At fixed frequency $\omega$ (with $k_0 = \omega/c$), the local dispersion relation can be written in the quadratic form

$g^{ij}(x')\, k_i k_j = n_0^2 k_0^2$, where $k_i$ are the covariant components of the local wavevector and $n_0$ is the refractive index of the virtual reference medium (typically $n_0 = 1$). One consistent choice relating the metric to the TO constitutive response is

$g^{ij}(x') \propto (\frac{\Lambda\Lambda^T}{J})^{-1}_{ij}$, up to an overall scalar convention that does not affect the phase-space arguments in Sec. 3. The role of this metric is purely structural: it packages the anisotropic TO dispersion into a Hamiltonian constraint surface $H = 0$ and makes the canonical phase-space structure explicit.

### 2.3. Hamiltonian rays and canonical phase-space structure

In the geometrical-optics regime, rays follow from the Hamiltonian. The Hamiltonian formulation employed here follows directly from the local dispersion relation (eikonal limit) of impedance-matched transformation media, as established within the standard TO mapping framework [1–3] and widely used for Hamiltonian ray tracing in TO media [21–22].

$$H(x',k) = \frac{1}{2}[g^{ij}(x')\, k_i k_j - n_0^2 k_0^2] = 0$$

with evolution

$$\dot{x}'^{\,i} = \frac{\partial H}{\partial k_i}, \dot{k}_i = -\frac{\partial H}{\partial x'^{\,i}},$$

where the dot denotes differentiation with respect to an affine ray parameter.

The only fact we will use in Sec. 3 is the following standard consequence of Hamiltonian dynamics: the flow is symplectic in $(x',k)$, i.e., it preserves the canonical two-form $dx'^i \wedge dk_i$ and therefore preserves the phase-space volume element $d^3x'\, d^3k$.

### 2.4. Phase-space measure and radiometric interpretation

To connect the canonical phase-space structure to radiometric quantities, we interpret radiance as a phase-space density on the dispersion surface $H = 0$. Denote by $L(x',\hat{s})$ the radiance (power per unit projected area per unit solid angle) along direction $\hat{s}$. The optical power transported by a narrow bundle intersecting an area element $dA$ within a solid-angle element $d\Omega$ can be written as

$$dP = L(x',\hat{s})\, dA_{\perp}\, d\Omega, dA_{\perp} = \cos\theta\, dA.$$

In passive, lossless media without internal emission or gain, radiance is conserved along rays (no source term in the transport equation), and the product

$E = \int L\, dA_{\perp} d\Omega$ defines the étendue of the beam, which is invariant along the propagation. Moreover, the canonical phase-space volume element is preserved by the symplectic flow. These two ingredients, radiance advection along rays plus phase-space incompressibility, are exactly what will yield, in Sec. 3, the invariance of étendue and the strict upper bounds on intensity enhancement in passive TO devices.

## 3. Radiance, étendue, and phase–space invariants in TO

In this section, we derive strict analytical bounds on the maximum attainable intensity enhancement in passive transformation-optical media. Our goal is to identify a phase-space invariant that is simultaneously compatible with the optical metric induced by a TO mapping and directly translatable into limits on radiance and field enhancement.
The derivation proceeds in a structured manner. We first state the physical goal and assumptions underlying the analysis, then establish the relevant phase-space invariants arising from Hamiltonian ray dynamics and radiance conservation. These invariants are subsequently specialized to transformation-optical mappings, leading to a closed-form bound on intensity enhancement.

Throughout this section, we assume that the device operates in the geometrical-optics regime $\lambda$ much smaller than the characteristic length scale of the medium), that the response is linear, passive, and time-invariant, and that the TO mapping reduces to the identity outside the device region.

### 3.1. Goal and assumptions

In this section, we establish the conceptual and analytical foundations required to derive rigorous bounds on intensity enhancement in passive TO media. While TO enables extreme spatial compression through coordinate mappings, it is not a priori evident whether such compression can lead to arbitrarily large field intensities once fundamental phase–space constraints are taken into account. The purpose of this section is therefore to identify the minimal physical ingredients needed to formulate such constraints and to clarify how they will be used in the subsequent derivations.

Our analysis is carried out within the framework of Hamiltonian ray optics, which naturally emerges in the geometrical–optics (eikonal) limit of impedance–matched TO media. As discussed in Refs. [1–3, 21–22], the local dispersion relation induced by a TO mapping can be expressed in terms of an effective optical metric $g_{ij}(x)$.

Building on Sec. 2.3, we now apply the same Hamiltonian formulation in the virtual (not primed coordinates) space of the TO medium. The equations are not re-derived; they are simply evaluated in $(x,k)$, which is the appropriate phase space for radiometric constraints and the subsequent intensity bounds. In three spatial dimensions, the corresponding Hamiltonian reads

$$H(x,k) = \frac{1}{2}[g^{ij}(x)\, k_i k_j - n_0^2 k_0^2] = 0,$$

where $x = (x_1,x_2,x_3)$ and $k = (k_1,k_2,k_3)$ denote the canonical position and covariant wave–vector coordinates, respectively, $g^{ij}$ is the inverse optical metric, and $n_0 k_0 = \omega/c$ is a frequency–dependent constant. Ray trajectories are governed by the standard Hamilton equations,

$$\dot{x}_i = \frac{\partial H}{\partial k_i}, \dot{k}_i = -\frac{\partial H}{\partial x_i},$$ where the dot denotes differentiation with respect to a ray parameter $s$(e.g., optical path length or an affine parameter).

A central consequence of this Hamiltonian structure is the preservation of the phase–space volume element along the ray flow, as dictated by Liouville's theorem in geometrical optics. In passive, linear, and time–invariant media without internal scattering, this implies that the phase–space measure remains invariant. This phase–space incompressibility constitutes the only property of the Hamiltonian dynamics required for the subsequent analysis and forms the basis of classical radiometric invariants used in nonimaging optics and radiometry [11,12,14].

Equally important for TO is that the coordinate mapping $x' = f(x)$ induces a transformation on covectors by pullback,

$$k_i = \Lambda^j{}_i\, k'_j,$$ so that the combined transformation $(x,k) \mapsto (x',k')$ is **canonical**. In particular, its total Jacobian in phase space satisfies

$$J_x\, J_k = 1$$

where $J_x$ is the spatial Jacobian determinant and $J_k$ is the corresponding Jacobian determinant in momentum space.

In the context of transformation optics, the coordinate mapping between virtual and physical spaces,

$x'_i = f_i(x_1,x_2,x_3)$, induces a canonical transformation in phase space, $(x,k)\leftrightarrow(x',k')$, which preserves the same phase–space volume elementbetween virtual space and real (physical) space:

$d^3x'\, d^3k'. = d^3x\, d^3k$. The combination of Hamiltonian ray evolution and TO–induced canonical mappings therefore preserves the phase–space measure globally. This identity is the precise mathematical statement behind the intuitive rule: spatial compression is necessarily accompanied by reciprocal angular (momentum-space) expansion.

Figure 1 schematically illustrates this canonical phase-space mapping and will be used as the conceptual entry point for the analysis in Secs. 3.2–3.4, where phase–space incompressibility is connected to radiance conservation, étendue invariance, and ultimately to an analytical upper bound on intensity enhancement.

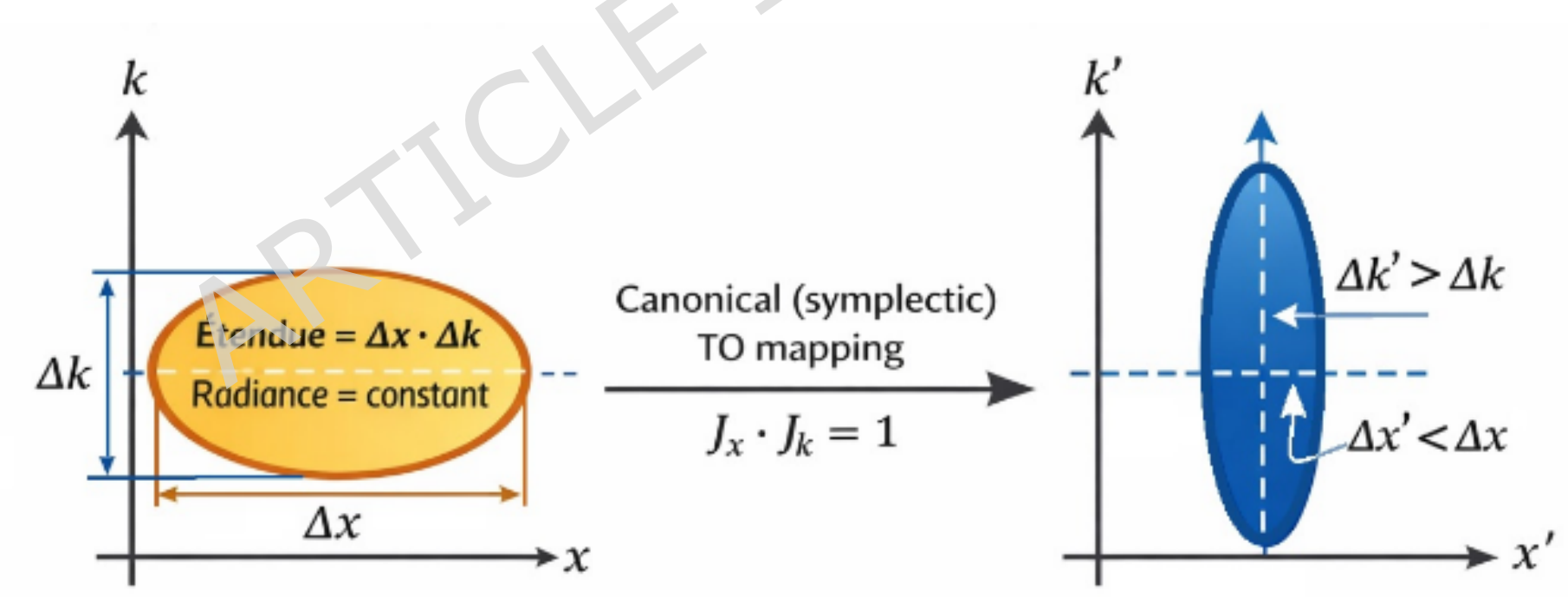

**Figure 1.** Canonical phase-space mapping induced by TO. A ray bundle occupying a finite volume in optical phase space $(x,k)$ in virtual space is mapped into physical space $(x',k')$ through a smooth, passive transformation-optical mapping. Spatial compression in real space is exactly compensated by reciprocal expansion in momentum (angular) space, such that the total phase-space volume (étendue) is preserved. This symplectic mapping satisfies $J_xJ_k = 1$ and provides a direct geometrical interpretation of Liouville's theorem in transformation-optical media.

### 3.2. Radiance as a phase–space density

To link this picture to radiometry, we need to connect radiance with a phase-space density. In geometrical optics, it is possible to define a distribution function $F(x,k)$ that qualitatively describes the amount of power carried by waves near the state $(x,k)$. A natural definition is to express the power carried by a small phase-space element as,

$$dP = C\,F(x,k)\,\delta(H(x,k))\,d^3x\,d^3k,$$

where $C$ is a normalization constant (depending on the field normalization) and $\delta(H)$ enforces the dispersion-surface constraint $H = 0$.

The radiance at a point $x$and along a given direction (or small solid angle $d\Omega$) is defined as the power per unit projected area and per unit solid angle,

$L(x,\hat{s}) = \dfrac{d^2P}{dA_\perp\,d\Omega}$,where $\hat{s}$is the propagation direction (e.g., the group velocity or, in the simplest limit, the direction of $k$), and $dA_\perp = \cos\theta\,dA$ is the effective area element normal to the ray direction.

To relate this to $F(x,k)$, note that the solid–angle element in $k$-space can be written as

$d\Omega = \dfrac{d^2k_\perp}{|k|^2}$,and the volume element in $k$-space as

$d^3k = |k|^2\,d|k|\,d\Omega$.On the dispersion surface $H = 0$, $|k|$ is approximately fixed, so we can perform the integral over $|k|$ and regard $F(x,k)$ as an effective density on $(x,\hat{s})$. In this framework, up to a constant factor, radiance can be taken proportional to this density:

$L(x,\hat{s}) \propto F(x,k(\hat{s}))$.If the medium is passive and contains no internal sources, the radiative transfer equation for unattenuated rays states that the radiance is conserved along each ray:

$\dfrac{d}{ds}L(x(s),\hat{s}(s)) = 0$(in the absence of emission, absorption, and scattering). This is equivalent to the statement that $F(x,k)$ is constant along the Hamiltonian flow on the dispersion surface $H = 0$:

$\dfrac{dF}{ds} = \dfrac{\partial F}{\partial x_i}\,\dot{x}_i + \dfrac{\partial F}{\partial k_i}\,\dot{k}_i = 0$.Taken together, “radiance is constant along rays” and “phase-space volume is preserved” express the essence of brightness and étendue in geometrical optics: compressing or expanding a beam can change the intensity $I$, but cannot increase the radiance $L$, because both $F$and the phase-space volume element $d^3 x\, d^3 k$ remain invariant.

This statement corresponds to the classical brightness (radiance) theorem, which forbids any increase of radiance in passive optical systems and underlies the conservation of étendue [11–14].

### 3.3. Étendue in the optical metric and its invariance under TO

Étendue (from the French word for "extent") measures the spread of a ray bundle in phase space. In the simplest spatially isotropic setting, for a beam crossing a surface $S$ of area $A$ within a solid-angle domain $\Omega$, the étendue is defined as

$$E = \int_S \int_\Omega n^2 \cos\theta \, dA \, d\Omega ,$$

where the factor $n^2$ accounts for the change in phase velocity in a medium of refractive index $n$, and $\cos\theta$ is the projection of the ray direction on the surface normal.

In the optical-metric formalism, this definition can be reexpressed using the relation between $dA$ and $d\Omega$ and the phase-space volume elements. Qualitatively, for a beam with radiance $L(x,\hat{s})$, the total power transmitted through a surface $S$ within a solid–angle domain $\Omega$ is

$$P = \int_S \int_\Omega L(x,\hat{s}) \cos\theta \, dA \, d\Omega .$$

If $L$ is approximately constant over that region, we can write

$$P \approx L_{max} E,$$

where $L_{max}$ is the maximum radiance in the beam, and

$$E = \int_S \int_\Omega \cos\theta \, dA \, d\Omega$$

appears as the étendue, up to conventions about where the factor $n^2$ is placed (either in $E$, in $L$, or absorbed into an effective solid angle). These definitional choices only modify overall constants and do not affect the structure of the final bounds.

Crucially, this étendue can be written equivalently as an integral over a subset of phase space:

$$E \propto \int_\Gamma d^3 x \, d^3 k ,$$

where $\Gamma$ is the set of phase-space points that constitute the ray bundle (i.e., all $(x,k)$ lying on the surface $S$ with the chosen directions). Because both the Hamiltonian ray flow and the TO mapping are canonical transformations, the volume $d^3 x\, d^3 k$ over $\Gamma$ is preserved along the rays. Therefore, for any other surface $\Sigma$ intersecting the same rays, the étendue has the same value:

$$E_{in} = E_{device} = E_{out}.$$

In other words, the TO mapping between virtual and physical spaces does not increase or decrease étendue; it only redistributes it among different coordinates. This is the metric-optics restatement of étendue conservation in the framework of TO.

### 3.4. Analytical bounds on intensity enhancement

We can now use this phase-space structure to derive analytical bounds on intensity and field enhancement. The basic idea is simple: if radiance in a passive device never exceeds its input value and étendue is conserved, then any increase in intensity must be compensated by a decrease in area or solid angle, not by an increase in brightness.

Before deriving the analytical bound, it is instructive to contrast the naive intuition commonly associated with transformation-optical concentrators with the correct physical behavior. A common misconception in the interpretation of transformation-optical concentrators is that strong spatial compression alone (purely geometric compression) implies unbounded intensity growth. As illustrated in Fig. 2(a), such a view neglects the accompanying redistribution in angular space. However, The correct physical picture, shown in Fig. 2(b), demonstrates that conservation of radiance and étendue enforces a strict upper bound on the attainable intensity enhancement.

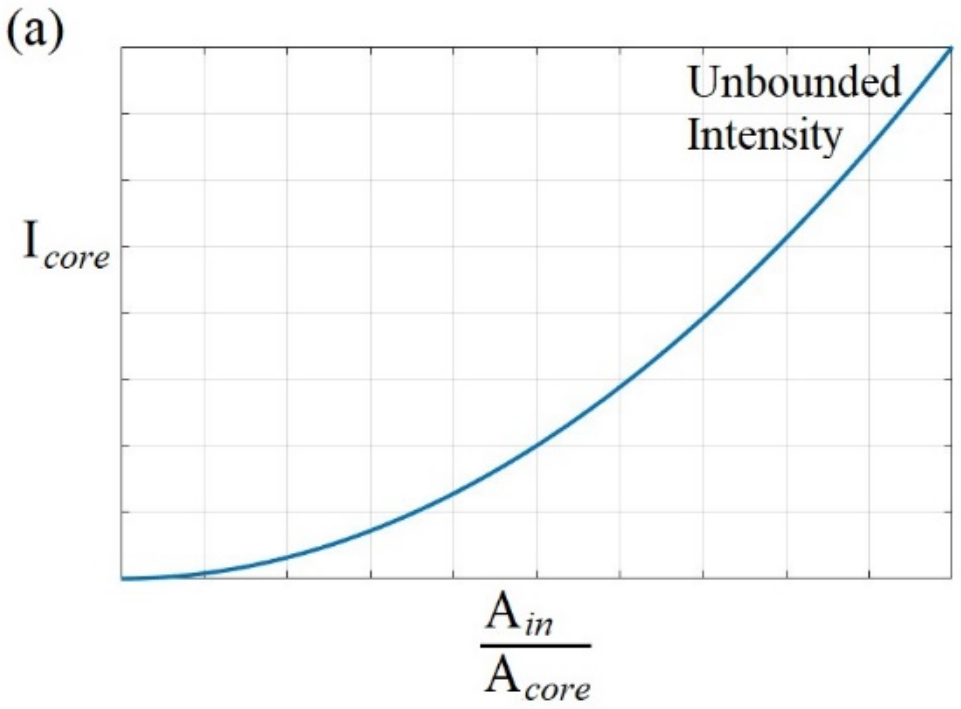

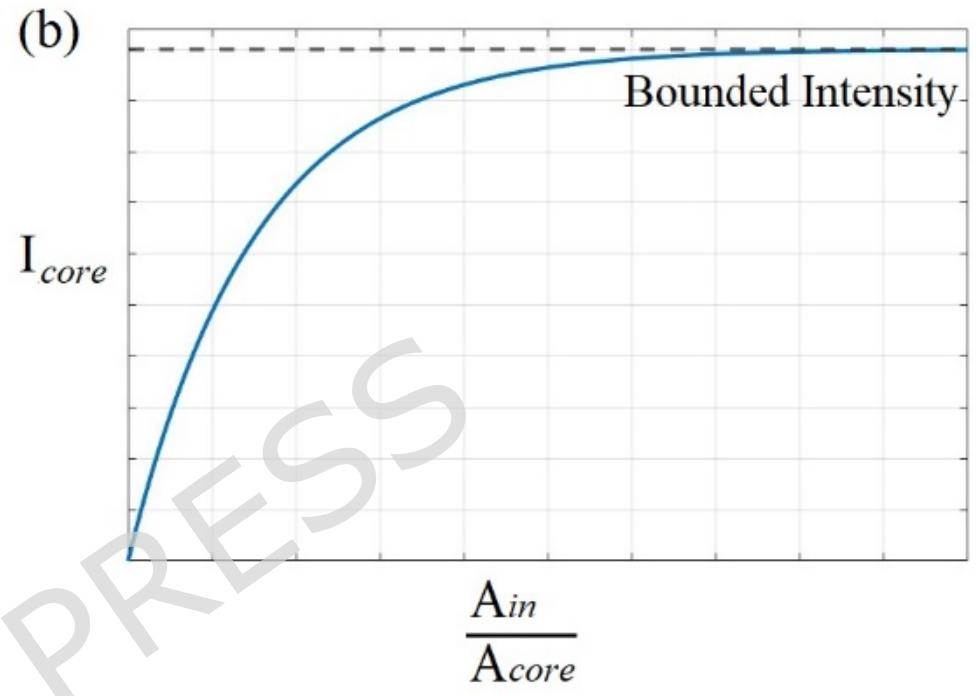

**Figure2.** Naive versus étendue-limited intensity enhancement. (a) Naive expectation predicts unbounded intensity: a common but incorrect intuition suggests that spatial compression alone leads to unbounded intensity growth. (b) The correct phase-space picture shows that angular spreading compensates spatial compression, enforcing a strict upper bound on average intensity enhancement determined solely by the area-compression ratio.

Suppose a TO device accepts a beam through an input surface $S_{in}$ of area $A_{in}$, within a solid–angle domain $\Omega_{in}$. Assume that the input radiance on this surface is bounded by $L(x,\hat{s}) \le L_{in,max}$(for example, a spatially extended, approximately uniform source with known maximum radiance). The input étendue is then approximately

$E_{in} \approx A_{in}\Omega_{in}$,if the beam is paraxial so that $\cos\theta \approx 1$, and variations of $L$ over the aperture are small.

Now consider a “core” region $\Sigma$ inside the device, e.g., the central cross section of a concentrator, with effective area $A_{core}$. By étendue conservation we must have

$$E_{core} = \int_{\Sigma}\int_{\Omega_{core}(x)} dA\, d\Omega = E_{in} \approx A_{in}\Omega_{in},$$

where

$\Omega_{core}(x)$ is the effective solid-angle domain of rays reaching the point $x \in \Sigma$. In the simple case where the device maps the input region uniformly onto the core (so that the TO mapping is approximately homogeneous over $\Sigma$), we may approximate $\Omega_{core}(x)$ by a nearly constant $\Omega_{core}$, which yields

$E_{core} \approx A_{core}\Omega_{core} = A_{in}\Omega_{in}$.From this relationship, the ratio of solid angles scales with the ratio of areas:

$\Omega_{core} \approx \Omega_{in}\dfrac{A_{in}}{A_{core}}$.The local intensity at a point $x$in the core can be written as

$$I_{core}(x) = \int_{\Omega_{core}(x)} L(x,\hat{s})\cos\theta\, d\Omega \le L_{in,max}\int_{\Omega_{core}(x)} \cos\theta\, d\Omega.$$

If the angles are small ($\cos\theta \approx 1$) and the radiance along the rays remains close to $L_{in,max}$(ideal, lossless case), this bound becomes approximately an equality:

$I_{core}(x) \approx L_{in,max}\,\Omega_{core}(x)$.Under the same uniformity approximation over $\Sigma$, the average intensity in the core region is

$$\grave{I}_{core} = \frac{1}{A_{core}}\int_{\Sigma} I_{core}(x)\, dA \approx L_{in,max}\,\Omega_{core} \approx L_{in,max}\,\Omega_{in}\frac{A_{in}}{A_{core}}.$$

Thus, the factor of average intensity increase in the core (relative to the average input intensity) is bounded by the area ratio:

$\dfrac{\grave{I}_{core}}{\grave{I}_{in}} \lesssim \dfrac{A_{in}}{A_{core}}$,where, for a simple uniform input,

$\grave{I}_{in} \approx L_{in,max}\,\Omega_{in}$.This inequality follows directly from two facts:

1. The radiance in a passive TO device cannot exceed $L_{in,max}$ along any ray (brightness conservation).
2. The étendue (effective phase–space volume) between the input surface and the core

region is conserved (étendue conservation).

The subtle point is that a TO mapping not only compresses the spatial area but also expands or contracts the associated solid angles in a reciprocal way: spatial compression with Jacobian $J_x = \det(\partial x/\partial X)$ is accompanied by an inverse Jacobian $J_k = J_x^{-1}$ in k-space, such that the total phase-space Jacobian satisfies $J_x J_k = 1$. As a result, the increase in intensity due to area compression is exactly balanced by a reduction in solid angle as seen from the source, so that brightness does not increase.

In practical designs, imperfections such as loss, scattering, strong anisotropy, and wave effects (diffraction) may cause the local intensity at specific points to be lower than the geometrical–optics prediction. However, under the assumptions of this paper, the inequalities above provide hard analytical bounds on field enhancement achievable by any passive TO device. In Sec. 4 we apply these bounds explicitly to several representative devices, including classical TO concentrators, zero index and optical null structures, and illusion media, and show which designs approach saturation of the bounds, and where diffraction, loss, or gain would be required to apparently "beat" the geometrical limit.

## 4. Applications to representative TO devices

In this section we apply the analytical framework of Sec. 3 to several canonical classes of TO devices. The purpose is twofold: (i) to show explicitly how radiance and étendue conservation constrain field enhancement in particular designs, and (ii) to clarify which architectures are closest to saturating the bounds and which are intrinsically far from optimal. We consider three representative families:

1. **Ideal TO concentrators** based on smooth radial mappings [4-6].
2. **Extreme–index platforms**, in particular zero index and optical null media (ZIM/ONM) [8-10,20] in supplementary appendix A.
3. **Illusion devices**, including cylindrical transformation media and optical null based illusions [7,10,17-19] in supplementary appendix B.

Throughout this section, we work with the assumptions stated in Section 3: linear, passive, time-invariant media that operate in the geometrical-optics regime, with a TO mapping that reduces to the identity outside of a finite device region.

### 4.1. Ideal TO concentrators: area compression and maximal intensity gain

We begin with an archetypal transformation optical concentrator in two or three dimensions. In virtual space, consider a homogeneous reference medium (typically vacuum) and an annular region

$R_1 < r < R_2$ that supports an incident beam of finite numerical aperture. A standard TO concentrator introduces a smooth, monotonic radial compression

$r' = f(r)$, mapping the virtual annulus onto a smaller physical region containing a core of radius $R_{core}$. In the ideal TO limit, the mapping is chosen such that the outer boundary is fixed, $f(R_2) = R_2$, while the inner boundary is compressed, for example $f(R_1) = \varepsilon R_1$ with $0 < \varepsilon \ll 1$[4–6].

#### 4.1.1. Spatial Jacobian and phase–space mapping

The spatial Jacobian associated with the radial transformation is

$J_x(r) = \det(\frac{\partial x'}{\partial x})$, which, in cylindrical or spherical symmetry, can be expressed in terms of $f(r)$ and its derivative. For a purely radial compression, the core region satisfies $|J_x| \ll 1$, reflecting the strong reduction of physical area (or volume) in that region.

As discussed in Sec. 3.1, the induced transformation on phase space is canonical: the momentum-space Jacobian $J_k$ is the inverse of the spatial one, so that the total phase-space Jacobian satisfies $J_x J_k = 1$. Consequently, any reduction of physical area in the core is accompanied by a reciprocal expansion in the associated solid-angle domain of rays that populate that region. In terms of étendue, we can write

$E_{in} \simeq A_{in}\,\Omega_{in}$ for the input aperture and

$E_{core} \simeq A_{core}\,\Omega_{core}$ for the core region, with

$E_{core} = E_{in}$ by étendue conservation (Sec. 3.3). Here $A_{in}$ is the illuminated area on the input surface, $A_{core}$ is the effective area of the core cross section, and $\Omega_{in}$, $\Omega_{core}$ are the corresponding solid-angle domains.

In the idealized case where the TO mapping distributes the incoming rays uniformly over the core region, the conservation law reduces to

$A_{core}\,\Omega_{core} = A_{in}\,\Omega_{in}$, so that

$\Omega_{core} \approx \Omega_{in}\,\dfrac{A_{in}}{A_{core}}$.

#### 4.1.2. Maximum achievable intensity enhancement

Let the input radiance on the entrance aperture satisfy

$L(x,\hat{s}) \le L_{in,max}$ for all relevant positions $x \in S_{in}$ and directions $\hat{s} \in \Omega_{in}$. As established in Sec. 3.2, radiance is conserved along rays in a passive TO device, so that

$L_{core}(x,\hat{s}) \le L_{in,max}$ everywhere in the core. The local intensity at a point $x$ within the core is

$$I_{core}(x) = \int_{\Omega_{core}(x)} L(x,\hat{s})\cos\theta\,d\Omega \le L_{in,max}\int_{\Omega_{core}(x)}\cos\theta\,d\Omega,$$

where $\theta$ is the angle between the ray direction and the local normal to a cross-sectional surface. In a paraxial regime, $\cos\theta \simeq 1$, and for a nearly uniform distribution of radiance over $\Omega_{core}(x)$ we obtain

$I_{core}(x) \approx L_{in,max}\,\Omega_{core}(x)$. Averaging over the core cross-section,

$$\bar{I}_{core} = \frac{1}{A_{core}}\int_{\Sigma} I_{core}(x)\,dA \approx L_{in,max}\,\Omega_{core},$$

and substituting $\Omega_{core} \approx \Omega_{in}A_{in}/A_{core}$ gives

$\bar{I}_{core} \approx L_{in,max}\,\Omega_{in}\,\dfrac{A_{in}}{A_{core}}$. For a uniform input source with

$\bar{I}_{in} \approx L_{in,max}\,\Omega_{in}$, the average intensity enhancement factor is bounded by

$\dfrac{\bar{I}_{core}}{\bar{I}_{in}} \lesssim \dfrac{A_{in}}{A_{core}}$. This result makes explicit the intuitive statement that an ideal TO concentrator can at most amplify intensity by the **area-compression ratio**, and that this enhancement is achieved without any increase in radiance. In the limit of a singular mapping with $A_{core} \to 0$, the intensity can diverge formally, but the radiance at every phase-space point remains bounded by $L_{in,max}$, in full agreement with the brightness theorems of nonimaging optics [11–13].

Realistic realizations of TO concentrators, including layered approximations and reduced–parameter designs [4–6], fall short of this ideal bound due to material dispersion, anisotropy truncation, and wave–optical effects. Nevertheless, the area ratio $A_{in}/$ $A_{core}$provides a rigorous analytical ceiling on what any passive TO concentrator can achieve in the geometrical–optics limit.

To validate the analytical bound derived in Sec. 3, we consider a radially symmetric transformation-optics concentrator and evaluate the maximum achievable **average intensity enhancement** as a function of the compression parameter $\varepsilon = f(R_1)/R_1$.

For each value of $\varepsilon$, the input area $A_{in}$and the compressed core area $A_{core}$are computed directly from the Jacobian of the radial coordinate transformation. The corresponding upper bound on the average intensity enhancement,$\frac{\grave{I}_{core}}{\grave{I}_{in}} \leq \frac{A_{in}}{A_{core}}$ is then evaluated without introducing any material-specific assumptions or full-wave numerical simulations. MATLAB calculation directly implements this relation for different values of the compression factor ε, without invoking any material-specific or full-wave numerical simulations. Figure 3 shows the resulting curve obtained by direct numerical evaluation of this closed-form analytical bound for different values of $\varepsilon$, derived in Sec. 3, where the maximum average intensity enhancement is given solely by the area-compression ratio $A_{in}/A_{core}$.

The solid line represents the étendue-limited enhancement, while the dashed reference line indicates the expected power-law scaling $\varepsilon^{-2}$for a two-dimensional concentrator. No fitting parameters, optimization procedures, or numerical field solvers are involved; the calculation strictly implements the analytical expressions derived in Sec. 3.

The numerical results demonstrate excellent agreement with the theoretical prediction: over the entire range $\varepsilon \in [10^{-2},1]$, the ratio , $\frac{\grave{I}_{core}}{\grave{I}_{in}}$ is indistinguishable from $A_{in}/A_{core}$ within numerical precision ( $< 10^{-11}$). This confirms that the enhancement is governed solely by étendue conservation and that no additional focusing or brightness increase occurs inside a passive TO concentrator.

Overall, Fig. 3 provides a direct numerical validation of the central theoretical result of this work: in an ideal, passive, lossless TO concentrator, the maximum achievable average intensity enhancement is exactly equal to the geometric area-compression ratio and cannot be exceeded.

Beyond the concentrator analyzed in Sec. 4.1, additional representative TO devices, including zero index/optical null media and illusion-based architectures, are presented in supplementary materials for completeness.

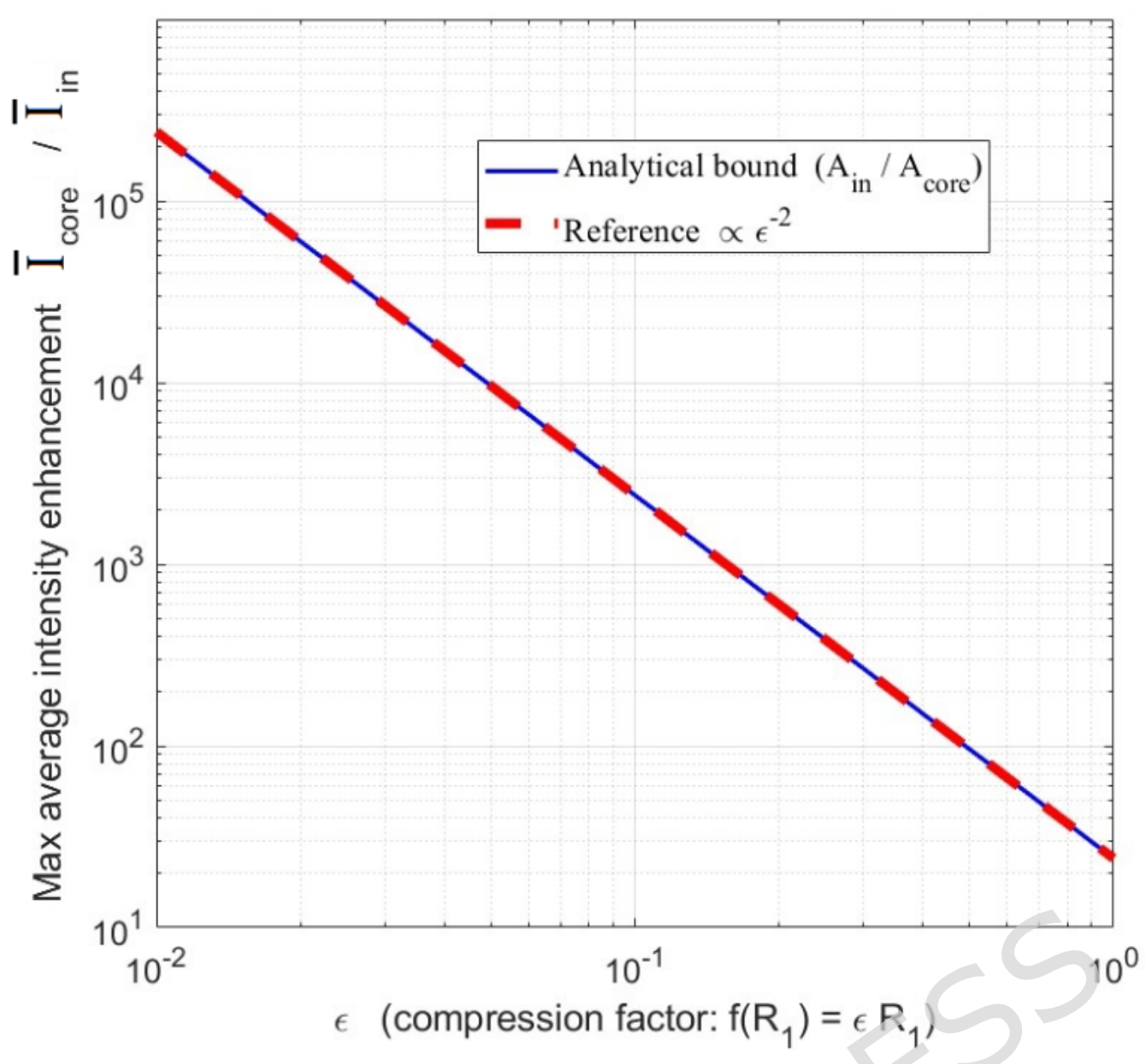

**Figure 3.** Maximum average intensity enhancement $\langle I_{core}\rangle/\langle I_{in}\rangle$ (Radiance/étendue bound ) as a function of the compression factor ε for an ideal TO concentrator. The data points are obtained by direct numerical evaluation of the analytical bound $\langle I_{core}\rangle/\langle I_{in}\rangle = A_{in}/A_{core} \propto \varepsilon^{-2}$, derived from radiance and étendue conservation. No full-wave simulations are involved; the curve represents the exact geometrical-optics limit imposed by phase-space invariance.

## 5. Discussion

### 5.1 phase-space origin of TO enhancement limits and scope of validity

We agree that a proportional scaling between attainable intensity enhancement and geometric compression has often been implicitly recognized in TO concentrator designs, especially for idealized radial mappings. The central contribution of the present work is not to restate that heuristic scaling, but to identify its fundamental origin and to elevate it to a well-defined, phase-space-based constraint within the assumption specified in this work by explicitly connecting TO to the classical brightness theorem through optical phase-space invariants. In particular, we show that any smooth, passive, impedance-matched TO mapping induces a canonical (symplectic) transformation on optical phase space. When combined with Hamiltonian ray dynamics in the associated optical metric, this yields a Liouville-type statement for TO: conservation of the appropriate phase-space measure, strict invariance (within the geometrical-optics regime) of étendue, and radiance conservation along rays. Within this framework, the familiar area-compression scaling of intensity enhancement is no longer a device-specific inference from field patterns or material tensors; instead, it emerges as a necessary consequence of phase-space incompressibility and brightness conservation. This resolves a recurrent ambiguity in the TO literature, namely, that very large local intensities observed in compressed regions can be misinterpreted as "brightness enhancement," whereas in a passive system they must be understood as intensity redistribution accompanied by a reciprocal modification of the angular (directional) distribution such that radiance does not increase.

This phase-space viewpoint also sharpens what is, and is not, meant by "arbitrarily large concentration" in passive TO. In the geometrical-optics limit, singular mappings can formally drive the core area toward zero, suggesting unbounded intensities. The present results clarify that such divergences are meaningful only as an asymptotic statement about geometric compression: they correspond to redistributing a fixed input étendue into progressively smaller spatial support, while the corresponding angular spread adjusts so that the overall phase-space content remains unchanged and radiance remains bounded by its input maximum. Stated differently, passive TO can reshape where and how power flows in space, but it cannot create additional phase-space density (brightness). This conceptual unification with the brightness theorem is the main novelty: it provides a rigorous bridge between the metric/canonical structure of TO and the radiometric invariants that govern all passive concentrators, thereby placing TO-based enhancement claims on the same footing as nonimaging optics.

### 5.2 Scope and Limitations of the Present Analysis

The bounds derived in this work rely explicitly on several physical assumptions. First, the analysis is carried out in the geometrical-optics (eikonal) regime, where the wavelength is much smaller than the characteristic length scale of the transformation medium, and a Hamiltonian ray description is valid. When transformation features approach the wavelength scale, diffraction and interference effects become significant, and the ray-based Liouville framework no longer provides a complete description. In such cases, wave effects typically enforce finite spot sizes and reduce achievable enhancement relative to the geometrical bound.

Second, the results assume linear, passive, time-invariant media without internal gain. Radiance conservation and étendue invariance follow from Hamiltonian flow only under these passive conditions. Incorporation of optical gain, non-Hermitian physics, or active modulation would take the system outside the scope of the present derivation and may modify the effective phase-space balance.

Third, the derivation presumes smooth, bijective coordinate transformations and impedance-matched implementations. Reduced-parameter realizations, impedance mismatch, strong material dispersion, absorption, or scattering introduce reflections and attenuation, which generally lower the attainable intensity relative to the idealized bound but do not increase radiance within passive settings.

Finally, the formulation applies to propagating (far-field) components for which radiance and solid-angle concepts are well defined. Near-field and evanescent regimes require a different framework, as conventional radiometric quantities are not directly applicable there.

Taken together, these assumptions delineate the precise domain in which the derived phase-space bounds apply. Within this domain, intensity enhancement in passive transformation-optical systems is fully constrained by étendue conservation and radiance invariance, whereas departures from these assumptions correspond to physically distinct regimes beyond the present analytical framework.

These distinctions clarify under which physical conditions extreme-index platforms (e.g., ZIM/ONM and illusion media) may approach the geometrical bound, while emphasizing that apparent “super-enhancement” must either reflect redistribution within conserved phase space or correspond to operation beyond the passive geometrical-optics regime considered here.

## 6. Conclusion

We have demonstrated that any passive TO devices, irrespective of its geometry, metric singularity, or material anisotropy, is fundamentally limited by the invariants of radiance and étendue. We demonstrated that the phase-space volume, and hence étendue, is precisely conserved by representing TO mappings as canonical transformations in optical phase space. Combined with radiance conservation along Hamiltonian rays, this yields strict upper bounds on achievable intensity and field enhancement.

For canonical TO concentrators, the maximum average intensity gain cannot exceed the geometric area-compression ratio $A_{in}/A_{core}$.

Zero-index and optical-null media, despite their extreme metrics, approach, but never surpass, this limit. Illusion devices redistribute rays without modifying radiance, and therefore cannot generate brightness beyond that of the source.

These findings establish a unified, strict set of framework for evaluating any TO design. They clarify which enhancements are physically achievable in passive systems and identify where additional physics, loss, gain, resonance, or full-wave interference, is required to exceed geometrical-optics expectations.

## Author Contributions Statement

**M.M.S.** developed the theoretical framework, derived the radiance-étendue bounds, and wrote the main manuscript text. **M.M.S. and M.S** contributed to the analytical formulation of TO mappings and provided critical revisions. **M.M.S** implemented the MATLAB simulations, generated the numerical figures, and assisted in the interpretation of the results. All authors reviewed and approved the final manuscript.

## Data Availability Statement

No datasets were generated or analyzed in this study. The work is purely analytical, and the numerical examples are illustrative and reproducible from the equations provided in the manuscript. MATLAB scripts are available from the corresponding author upon reasonable request.

## Funding Declaration

No funding

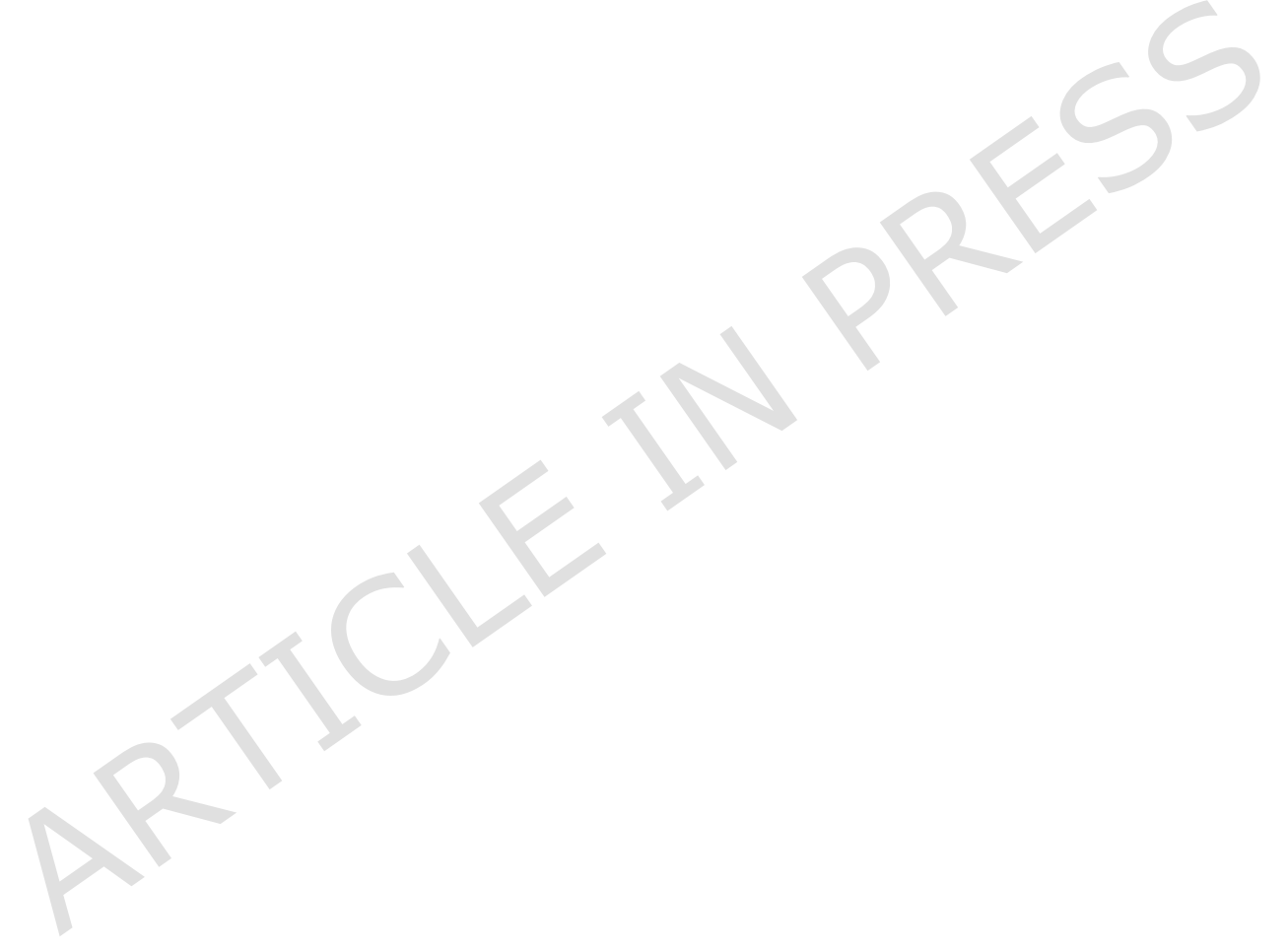